\documentclass[aps,11pt,nofootinbib,preprintnumbers,groupedaddress,superscriptaddress]{revtex4}
\usepackage{amssymb, amsfonts, amsmath, bm}
\usepackage[
]{graphicx}
\usepackage{color}

\usepackage{mathrsfs}
\usepackage{enumerate}

\usepackage[
]{hyperref}
\hypersetup{%
 setpagesize=false,
 bookmarksnumbered=true,%
 bookmarksopen=true,%
 colorlinks=true,%
 linkcolor=blue,%
 citecolor=red}

\begin{document}
\title{
Particle dynamics in the Newtonian potential 
sourced by a homogeneous circular ring 
}
\author{{\large 
Takahisa Igata}}
\email{igata@post.kek.jp}
\affiliation{KEK Theory Center, 
Institute of Particle and Nuclear Studies, 
High Energy Accelerator Research Organization, Tsukuba 305-0801, Japan}
\date{\today}
\preprint{KEK-Cosmo-253, KEK-TH-2213}

\begin{abstract}
Newtonian gravitational potential sourced by a homogeneous circular ring in arbitrary dimensional Euclidean space takes a simple form if the spatial dimension is even. In contrast, if the spatial dimension is odd, it is given in a form that includes complete elliptic integrals.
In this paper, we analyze the dynamics of a freely falling massive particle in its Newtonian potential. 
Focusing on circular orbits on the symmetric plane where the ring is placed, we observe that they are unstable in 4D space and above, while they are stable in 3D space. The sequence of stable circular orbits disappears at $1.6095\cdots$ times the radius of the ring, which corresponds to the innermost stable circular orbit (ISCO). On the axis of symmetry of the ring, there are no circular orbits in 3D space but more than in 4D space. 
In particular, the circular orbits are stable between the ISCO and infinity in 4D space and between the ISCO and the outermost stable circular orbit in 5D space. 
There exist no stable circular orbits in 6D space and above. 
\end{abstract}
\maketitle

\section{Introduction}
\label{sec:1}
Physics in higher-dimensional spacetimes 
is a target of research attracting much attention 
from various fields. 
In high energy physics, 
higher-dimensional spacetimes inevitably 
appear in the process of seeking 
a unified theory of force.
For example, the idea of the AdS/CFT correspondence~\cite{Maldacena:1997re, Gubser:1998bc, Witten:1998qj, Witten:1998zw} strongly motivates the study of higher-dimensional classical gravitational theories because we can calculate various physical quantities in quantum field theories on the boundary side using classical bulk gravity. 
Furthermore, the study of higher-dimensional spacetimes implies parametrization of the spacetime dimensions~\cite{Emparan:2013moa}. Through this, we can understand the dimensional dependence of classical gravity, 
e.g., the particularity of physical phenomena in given dimensions.

In this context, the study of high-dimensional black holes has made great progress~\cite{Emparan:2008eg, Tomizawa:2011mc}. 
One of the breakthroughs was the discovery of the black ring solution in 5D spacetime~\cite{Emparan:2001wn}.
The existence of this novel solution is symbolic of the diversity of high-dimensional classical gravity. After this discovery, many black ring solutions in 5D have been 
discovered~\cite{Pomeransky:2006bd, Mishima:2005id, Elvang:2007rd, 
Iguchi:2007is, Evslin:2007fv, Izumi:2007qx, Elvang:2007hs}. 
The analytical solutions of a black ring are expected to be found even in spacetimes more than 6D; however, so far they have not been found. There are some efforts to find them by numerical integrations~\cite{Kleihaus:2012xh, Dias:2014cia}. 
On the other hand, there also several analytical approaches to construct higher-dimensional black rings such as the blackfold method \cite{Emparan:2007wm,Armas:2014bia} or 
the large spacetime dimension limit~\cite{Tanabe:2015hda, Tanabe:2016pjr}.

If exact solutions to black rings and other novel high-dimensional black objects are available, we can understand these physics, e.g., through geodesic analysis~\cite{Nozawa:2005eu, Hoskisson:2007zk, Igata:2010ye, Igata:2010cd, Grunau:2012ai, Igata:2013be, Diemer:2014lba, Igata:2014xca, Igata:2014bga, Tomizawa:2019egx}.
However, even if an exact solution has not yet been found, we can discuss some fundamental properties of black objects because the distant gravitational field of a black object is approximated by the Newtonian gravitational potential sourced by a corresponding source. 
In fact, this property is utilized in the blackfold approach as mentioned above~\cite{Emparan:2009at}. From the perspective of geodesic analysis of black objects, 
the motion of particles there can be viewed as timelike geodesics in the weak gravity region. 
Therefore, once the gravitational field is obtained, even in an approximate form, we can read some properties from the motion of probe particles. 
Furthermore, if the gravitational field contains the spacetime dimensions as a parameter, then the dimensionality of gravity appears in the particle dynamics. In particular, the dimensionality can be expected to appear clearly in the existence of stable circular orbits (or stable bound orbits), which are the most fundamental orbits of black object spacetimes.

The aim of this paper is to reveal the dimensionality of 
the gravitational field of black rings through the dynamics of a test particle as a probe.
However, since the black ring solution has not yet been parametrized in a spacetime dimension,  we consider the dynamics of particles in the Newtonian gravitational potential sourced by a homogeneous circular ring source, which is parametrized by the spacetime dimensions.
Focusing on the existence of stable circular orbits of a 
massive particle around the ring source, 
we make sure the dimensionality appears 
for the particle motion.

The dynamics in the Newtonian gravitational potential are also important in the context of the integrability of the geodesic equations and spacetime hidden symmetry.
There are some examples where the integrable property of a particle mechanical system is restored in the Newtonian limit.
We can find such an example in the relation between 
particle mechanics in a multi-black hole spacetime and 
Euler's three body problem---particle motion around two fixed center placed on 3D Euclidean space is known as integrable. 
We can find another example in massive particle mechanics
in 5D black ring spacetime. 
In general, 
a massive particle behaves chaotic (i.e., nonintegrable) 
around the black ring~\cite{Igata:2010cd}; however, 
the integrable property is restored in the Newtonian limit, 
which is identified with a particle mechanics in 
a ring source potential.%
\footnote{The integrability of the equation of particle motion 
in a ring source potential in 4D Euclidean space is closely related to the 
integrability of Euler's three body problem~\cite{Igata:2014bga}. }

This paper is organized as follows. In the following section, 
we derive the Newtonian gravitational potential 
sourced by a homogeneous circular ring source 
in arbitrary dimensional Euclidean space.
Furthermore, we show that the potential reduces to a simple form in the case that the spatial dimension is even. 
In Sec.~\ref{sec:3}, 
we focus on particle mechanics in the Newtonian potential and 
consider the existence of stable circular orbits and its dependence of spatial dimensions. 
Section~\ref{sec:4} is devoted to a summary and discussions.

\section{Newtonian potential sourced by a homogeneous circular ring in $\mathbb{E}^n$}
\label{sec:2}
We derive the Newtonian gravitational potential sourced by a homogeneous 
circular ring in $n$-dimensional Euclidean space~$\mathbb{E}^n$. 
Let $(\zeta, \psi)$ be polar coordinates on a 2D plane in $\mathbb{E}^n$ and 
$(\rho, \phi_1, \ldots, \phi_{n-3})$ be spherical coordinates on 
the $(n-2)$-dimensional space perpendicular to it, where $n\geq3$. 
The Euclidean metric in these coordinates is given by
\begin{align}
\mathrm{d\ell}^2=\mathrm{d}\zeta^2+\zeta^2\:\! \mathrm{d}\psi^2+ \mathrm{d}\rho^2+\rho^2 \mathrm{d}\Omega_{n-3}^2,
\end{align}
where $\mathrm{d} \Omega_{n-3}^2$ is the metric on the unit $(n-3)$-sphere. 
Let us focus on a homogeneous ring-shaped gravitational source with radius $R$. 
Without loss of generality, 
the mass density function of the ring is given by 
\begin{align}
\sigma(\bm{r})=\frac{M}{2\pi \Omega_{n-3}\:\!\zeta\:\!\rho^{n-3}}\:\!\delta(\zeta-R)\:\!\delta(\rho),
\label{eq:sigma}
\end{align}
where $M$ is the total mass of the ring, and $\Omega_{n-3}$ is the area of unit $(n-3)$-sphere, and $\delta(\cdot)$ is the delta function.
Given the distribution of sources $\sigma$, we can determine Newtonian potential $\Phi_n(\bm{r})$ by solving the field equation of the $n$-dimensional Newtonian gravity,
\begin{align}
\label{eq:feq}
\nabla^2\Phi_n(\bm{r})=\Omega_{n-1}G\:\! \sigma(\bm{r}), 
\end{align}
where $\nabla^2$ denotes the Laplacian of $\mathbb{E}^n$, and $G$ is the gravitational constant.%
\footnote{The physical dimension of the gravitational constant $G$ in ordinary units is 
$[\:\!G\:\!]=(\mathrm{length})^n(\mathrm{mass})^{-1}(\mathrm{time})^{-2}$.}

Now, assuming the matter distribution~\eqref{eq:sigma}, we solve the field equation~\eqref{eq:feq}. Using the Green function, the formal solution can be expressed as
\begin{align}
\Phi_n(\bm{r})=-\frac{G}{n-2}\int_{\mathbb{E}^n}\mathrm{d}^nr'\frac{\sigma(\bm{r}')}{|\bm{r}-\bm{r}'|^{n-2}}.
\label{eq:phi}
\end{align}
Integrating with respect to the coordinates other than $\psi'$, 
we obtain
\begin{align}
\Phi_n(\bm{r})
&=-\frac{GM}{2(n-2)\pi}\int_0^{2\pi}\frac{\mathrm{d}\psi'}{\left[\zeta^2+\rho^2+R^2-2R\:\! \zeta \cos \psi'\right]^{(n-2)/2}}
\\
&=-\frac{GM}{2(n-2)\pi \:\!r_+^{n-2}} \int_0^{2\pi}\mathrm{d}\psi'\left[\:\!
1-z\:\! \cos^2 \frac{\psi'}{2}\:\!\right]^{(2-n)/2},
\label{eq:phi_psi}
\end{align}
where 
\begin{align}
&r_\pm=\sqrt{(\zeta \pm R)^2+\rho^2},
\\
\label{eq:z}
&z=\frac{4R\:\!\zeta}{r_+^2}=1-\frac{r_-^2}{r_+^2}. 
\end{align}
It is convenient to make the change of variables, $u=\cos^2 \psi'/2$, in terms of which Eq.~\eqref{eq:phi_psi} becomes
\begin{align}
\label{eq:Euler}
\Phi_n(\bm{r})
&=-
\frac{GM}{(n-2)\pi r_+^{n-2}} \int_0^1 u^{a-1}(1-u)^{c-a-1}(1-z\:\! u)^{-b}\:\!\mathrm{d}u,
\end{align}
where we have specified constants $a$, $b$, and $c$ by
\begin{align}
&a=\frac12, \quad
b=\frac{n-2}{2},
\quad
c=1,
\end{align}
respectively. Using Euler's integral representation of the Gaussian hypergeometric function $F(a, b, c; z)$,%
\footnote{The Gaussian hypergeometric function is defined by
\begin{align}
F(a, b, c; z)=\frac{\Gamma(c)}{\Gamma(a)\Gamma(c-a)}\int_0^1 u^{a-1}(1-u)^{c-a-1}(1-zu)^{-b}\mathrm{d}u,
\end{align}
where $\Gamma(\cdot)$ denotes the Gamma function. 
}
we can represent the right-hand side of Eq.~\eqref{eq:Euler} as
\begin{align}
\label{eq:Phi}
\Phi_n(\bm{r})=-\frac{GM}{(n-2) \:\!r_+^{n-2}}\:\!F\left(\frac12, \frac{n-2}{2},1; 1-\frac{r_-^2}{r_+^2}\right).
\end{align}
This is the expression of the Newtonian gravitational potential sourced 
by a homogeneous circular ring in $\mathbb{E}^n$.

It is worthwhile to present a more explicit form of $\Phi_n(\bm{r})$ in each dimensions to understand the nature of the gravitational field and the behavior of test particles. Moreover, the explicit form may have some implications for the discovery of higher-dimensional black ring solutions because the weak-gravity limit of a time-time metric component of a black ring is expected to include $\Phi_n(\bm{r})$. In what follows, analyzing the dependence of the hypergeometric function in Eq.~\eqref{eq:Phi} on the parity of $n$, 
we clarify the simple mathematical structure of $\Phi_n(\bm{r})$ for even $n$ (see also Refs.~\cite{Lunin:2002iz,Emparan:2007wm}). 

In the following, we restrict our attention to the case where $n$ is even, i.e., 
\begin{align}
n=2m,
\end{align}
where $m$ is an integer greater than or equal to $2$.
For simplicity, we introduce a new variable
\begin{align}
\chi=\frac{r_-}{r_+}.
\end{align}
Then the hypergeometric function appearing in Eq.~\eqref{eq:Phi} is rewritten as
\begin{align}
F(a,b,c; 1-\chi^2)
&=\frac{\Gamma(c)\,\Gamma(c-a-b)}{\Gamma(c-a)\,\Gamma(c-b)}\,F(a,b,a+b+1-c; \chi^2)
\cr
&\quad +\frac{\Gamma(c)\,\Gamma(a+b-c)}{\Gamma(a)\,\Gamma(b)}\,\chi^{2(c-a-b)}\,F(c-a,c-b,1+c-a-b; \chi^2)
\\
\label{eq:secondline}
&=\frac{\Gamma(1)\,\Gamma(m-3/2)}{\Gamma(1/2)\,\Gamma(m-1)}\,\chi^{3-2m}\,F\left(\frac12, 2-m, \frac52-m; \chi^2\right). 
\end{align}
Since the argument $2-m$ of $F$ in Eq.~\eqref{eq:secondline} is not positive, the power series is truncated at a finite term. 
Thus, the Newtonian potential $\Phi_n(\bm{r})$ for even $n$ is given in the following form:
\begin{align}
\Phi_{2m}(\bm{r})=-\frac{GM}{(m-1) (2\:\!r_+r_-)^{m-1}}\,
\sum_{\nu=0}^{m-2}\alpha_\nu\, \left(\frac{r_-}{r_+}\right)^{2\nu+2-m},
\label{eq:polynomial}
\end{align}
where 
\begin{align}
\alpha_\nu=\frac{(2\nu-1)!!}{\nu!} \frac{[\:\!2(m-\nu-2)-1\:\!]!!}{(m-\nu-2)!}.
\end{align}
Substituting $m=2, 3, 4$, and $5$ into Eq.~\eqref{eq:polynomial}, we obtain 
\begin{align}
\Phi_{4}(\bm{r})&=-\frac{GM}{2\:\!r_+r_-},
\\
\Phi_{6}(\bm{r})&=-\frac{GM}{8\:\! (r_+r_-)^2}\left(\chi^{-1}+\chi\right),
\\
\Phi_{8}(\bm{r})&=-\frac{GM}{16\:\!(r_+r_-)^3} \left(\chi^{-2}+\frac{2}{3}+\chi^2\right),
\\
\Phi_{10}(\bm{r})&=-\frac{5\:\!GM}{128\:\!(r_+r_-)^4} \left(\chi^{-3}+\frac{3}{5}\chi^{-1}+\frac{3}{5}\chi+\chi^3\right).
\end{align}
The form of $\Phi_{4}(\bm{r})$ appears in the weak-gravity limit of a time-time metric component of the Emparan-Reall black ring solution~\cite{Igata:2014bga}. 
Similarly, if there are black ring solutions with spatial dimensions greater than $5$, then the time-time component of these metrics should also include $\Phi_n(\bm{r})$ of Eq.~\eqref{eq:Phi} in the weak-gravity limit. 
Though $\Phi_n(\bm{r})$ takes the above simple form for even $n$ 
but is given in terms of complete elliptic integrals for odd $n$ as
\begin{align}
\Phi_{3}(\bm{r})&=-\frac{2GM}{\pi} \frac{K(z)}{r_+},
\\
\Phi_{5}(\bm{r})&=-\frac{2GM}{3\pi} \frac{E(z)}{r_+r_-^2},
\\
\Phi_{7}(\bm{r})&
=-\frac{2GM}{15 \pi \:\!r_-^5}\left[\:\!
-\chi^3 K(z)+2\chi (1+\chi^2)E(z)
\:\!\right],
\\
\Phi_{9}(\bm{r})&
=-\frac{2GM}{105\pi\:\!r_-^7}\left[\:\!
(8\chi+7\chi^3+8\chi^5)E(z)-4\chi^3(1+\chi^2)K(z)
\:\!\right],
\end{align}
where $K(z)$ is the complete elliptic integral of the first kind, and 
$E(z)$ is the complete elliptic integral of the second kind.%
\footnote{We adopt the following convention of the complete elliptic integrals:
\begin{align}
K(z)&=\int_0^{\pi/2}\frac{\mathrm{d}\theta}{\sqrt{1-z \sin^2\theta}},
\\
E(z)&=\int_0^{\pi/2} \sqrt{1-z \sin^2\theta}\:\!\mathrm{d}\theta.
\end{align}
} %
This fact implies that a black ring solution with even spatial dimensions has a simpler metric form than that with odd spatial dimensions.

\section{Dynamics of a massive particle in the ring source potential}
\label{sec:3}
We consider the dynamics of a freely falling massive particle in the Newtonian gravitational potential~\eqref{eq:Phi}. 
Let $m$ be the mass of a particle and 
 $p_i$ be canonical momenta conjugate to coordinates, $x^i$. 
Then the Hamiltonian of a massive particle is given by%
\footnote{
In the spheroidal coordinates,
\begin{align}
\zeta=R\:\!\xi\:\!\eta, \quad \rho=R\sqrt{(\xi^2-1)(1-\eta^2)}, 
\end{align}
or equivalently, 
\begin{align}
r_\pm=R \:\!(\xi\pm \eta), 
\end{align}
the separation of variables in the Hamilton-Jacobi equation 
for $H_n$ 
occurs for $n=4$~\cite{Igata:2014bga}. 
However, it is an open question whether there exists a coordinate system in which the separation of variables occurs for $n\geq 5$.}
\begin{align}
\label{eq:H}
H_n=\frac{1}{2m}\left[\:\!
p_\zeta^2+p_\rho^2+\frac{L^2}{\zeta^2}+\frac{Q^2}{\rho^2}
\:\!\right]+m \Phi_{n}(\bm{r}),
\end{align}
where $L$ and $Q^2$ are defined by, respectively, 
\begin{align}
L&=p_\psi, \\
Q^2&=\left\{
\begin{array}{ll}
0&(n=3)
\\
\gamma^{ab}p_a p_a&(n\geq4),
\end{array}
\right.
\end{align}
where $\gamma^{ab}$ is the inverse metric on the unit $(n-3)$-sphere, 
and $a$ and $b$ label $(\phi_1, \ldots, \phi_{n-3})$.
Note that $L$ is a constant of motion associated with axial symmetry, 
and also $Q$ is a constant of motion associated with 
spherical symmetry in the $(n-2)$-dimensional subspace spanned by $(\rho, \phi_1,\ldots, \phi_{n-3})$. 
Note that $Q$ must vanish for $n=3$ because the term proportional to $Q^2$ in $H_3$ does not exist. 
Furthermore, $H_n$ itself is also conserved because $\Phi_n$ is time independent. Hence, 
the energy equation becomes
\begin{align}
&\frac{m}{2}(\dot{\zeta}^2+\dot{\rho}^2)+V_n=\mathcal{E},
\\
\label{eq:Vn}
&V_n(\zeta, \rho)=\frac{L^2}{2\:\!m\:\!\zeta^2}+\frac{Q^2}{2\:\!m\:\!\rho^2}
+m\:\! \Phi_n(\bm{r}),
\end{align}
where $\mathcal{E}$ is a constant energy of a particle, and 
the dots are derivatives with respect to time. 
We refer to $V_n$ as the effective potential. 
The first two terms in $V_n$ are centrifugal potentials, which do not depend on $n$, unlike the third term. In what follows, we consider the dynamics of a massive particle moving through particular regions determined by the symmetry of $\Phi_n$.

\subsection{Dynamics on the symmetric plane $\rho=0$}
We focus on the motion of a particle constrained to the 2D plane where the ring source is located. If we launch a particle into this plane, the particle inevitably 
continue to move over it because the Hamiltonian $H_n$ is $\mathbb{Z}_2$ symmetric with respect to $ \rho=0$.
Therefore, we refer to the plane as the symmetric plane below. 
To move on the symmetric plane, a particle must have $Q=0$. 
Thus, the explicit form of Eq.~\eqref{eq:Vn} on the symmetric plane is given by
\begin{align}
\label{eq:Vncomplete}
V_n(\zeta)=\frac{L^2}{2m\zeta^2}-\frac{GMm}{(n-2) (\zeta+R)^{n-2}}F\left(
\frac12, \frac{n-2}{2}, 1; z
\right),
\end{align}
where we have defined $V_n(\zeta)=V_n(\zeta, 0)\big|_{Q=0}$ and use the notation throughout this subsection for abbreviation, and $z$ in Eq.~\eqref{eq:z} reduces to 
\begin{align}
z=\frac{4R\:\!\zeta}{(\zeta+R)^2}.
\end{align}
One of the important properties of particles moving on the symmetric plane is that they remain stably constrained to the plane even if small perturbations are applied perpendicularly to it. In fact, the expansion of $V_n(\zeta,\rho)\big|_{Q=0}$ around $\rho=0$,
\begin{align}
V_n(\zeta,\rho)\big|_{Q=0}=V_n(\zeta)+\frac{GMm}{2(\zeta+R)^n}\left[\:\!
F\left(\frac12, \frac{n-2}{2},1; z
\right)+\frac{2R\:\!\zeta}{(\zeta+R)^2}F\left(
\frac32, \frac{n}{2}, 2; z
\right)
\:\!\right]\rho^2+O(\rho^4),
\end{align}
shows that 
the effective potential makes a local minimum at $\rho=0$ because the coefficient of $\rho^{2}$ is positive.

First, let us consider the dynamics of a particle near the center of the ring. 
The expansion of 
$V_n(\zeta)$ around $\zeta=0$ is given by
\begin{align}
\label{eq:Vnsmallzeta}
V_n(\zeta)=\frac{L^2}{2\:\!m\:\!\zeta^2}-\frac{GMm}{(n-2) R^{n-2}}
-\frac{(n-2) GMm}{4R^n} \zeta^2-\cdots.
\end{align}
The qualitative behavior of particles near the center is independent on the spatial dimensions because the power of $\zeta$ does not depend on $n$.
If $L\neq0$, the centrifugal barrier appears near the center. 
If $L=0$, the function $V_n(\zeta)$ takes a local maximum at $\zeta=0$ 
because the coefficient of $\zeta^2$ in the third term is negative, which shows outward gravitational force from the ring.
Therefore, a particle can stay at the unstable 
equilibrium point and then has an energy $\mathcal{E}=-GMm/[(n-2)R^{n-2}]$, which is
consistent with the constant value of the second term of $V_n(\zeta)$.

Next, we consider $V_n(\zeta)$ in the asymptotic region, where 
$\zeta/R\gg 1$. The asymptotic expansion of $V_n(\zeta)$ is given by 
\begin{align}
\label{eq:Vnlargezeta}
V_n(\zeta)= \frac{L^2}{2\:\!m\:\!\zeta^2}
-\frac{GMm}{n-2} \frac{1}{\zeta^{n-2}}-\frac{(n-2)GMm}{4} \frac{R^2}{\zeta^n}
-\frac{n^2(n-2)GMm}{64}\frac{R^4}{\zeta^{n+2}}
-\cdots. 
\end{align}
The second term is the monopole term, and the higher-order terms after the third term include contribution from the ring shape. Unlike Eq.~\eqref{eq:Vnsmallzeta}, the power of $\zeta$ for all gravitational term depends on $n$. In what follows, we analyze the effect of $n$-dependent differences in the functional form of $V_n(\zeta)$ on the dynamics of particles.

Let us focus on the case where $n=3$. 
Though the black ring solution is forbidden by the uniqueness theorem of the black hole in $n=3$, some properties must be of great help to know the dimensionality of the particle dynamics around a ring source. Furthermore, in the context of astrophysics, such dynamics would be worth considering. The asymptotic expansion of $V_3(\zeta)$ takes the form,
\begin{align}
V_3(\zeta)=-\frac{GMm}{\zeta}+\frac{L^2}{2\:\!m\:\!\zeta^2}-\frac{GMm}{4} \frac{R^2}{\zeta^3}-\cdots. 
\end{align}
The leading order is the monopole term, and the subleading order is the centrifugal term. 
By the same mechanism as in the case of a point source, these two terms can form a potential well, so that particles can stably stay near its bottom.
This means that there exist stable circular orbits in the region far from the ring.

Do the stable circular orbits still exist in the vicinity of the ring source? To answer this question, we analyze conditions for the existence of stable circular orbits using the complete form of $V_3(\zeta)$.
The conditions for a particle moving on a circular orbit are given by 
\begin{align}
V_3'(\zeta)=0, \quad V_3(\zeta)=\mathcal{E}, 
\end{align}
where the prime denotes the derivative with respect to $\zeta$.
Solving these equations in terms of $\mathcal{E}$ and $L$, we obtain the energy and angular momentum of a particle moving in a circular orbit as functions of the orbital radius, respectively, 
\begin{align}
\mathcal{E}_{0}&=\frac{GMm}{2\pi} \left[\:\!
\frac{E(z)}{\zeta-R}-\frac{3 K(z)}{\zeta+R}
\:\!\right],
\\
L_{0}^2&=\frac{GMm^2}{\pi}\zeta^2 \left[\:\!
\frac{E(z)}{\zeta-R}+\frac{K(z)}{\zeta+R}
\:\!\right].
\end{align}
We observe that there are no circular orbits between the origin and the ring, $0<\zeta<R$, because 
 $L_{0}^2$ is negative there. 
On the other hand, we find circular orbits in $R<\zeta<\infty$ because $L_0^2$ is positive. Not all of these circular orbits are stable, and the stability of a circular orbit requires the additional condition,
\begin{align}
\label{eq:stablility}
V_3''(\zeta)|_{L=L_0}=\frac{2GMm}{\pi \zeta^2(\zeta+R)}\left[\:\!
K(z)-\frac{R^2}{(\zeta-R)^2} E(z)
\:\!\right]\geq 0.
\end{align}
This holds in the range,
\begin{align}
\zeta_0 \leq \zeta <\infty,
\end{align}
where we have determined the marginal value $\zeta_0$ by solving Eq.~\eqref{eq:stablility} numerically when the equality holds,
\begin{align}
\zeta_0/R=1.6095\cdots. 
\end{align}
After all, the sequence of stable circular orbits exists from infinity to the innermost radius $\zeta_0$, which is slightly outside the ring.%
\footnote{This value was obtained in Ref.~\cite{DAfonseca:2005acz} in the context of geodesic motion in the Bach-Weyl ring spacetime. }
The marginal circular orbit at $\zeta=\zeta_0$ is often referred to as 
the innermost stable circular orbit (ISCO), which appears as a feature of stable circular orbits in black hole spacetimes. The energy and angular momentum at the ISCO are $\mathcal{E}_0/(GMm/R)=-0.24641\cdots$ and $L_0^2/(GMm^2R)=2.3463\cdots$.

Let us consider the case where $n=4$. The motion of particles in $\Phi_4$ reflects the nature of gravity in the asymptotic region of the Emparan-Reall black ring spacetime. 
The effective potential~\eqref{eq:Vnlargezeta} for $n=4$ takes the form,
\begin{align}
V_4(\zeta)=\left(\frac{L^2}{2m}-\frac{GMm}{2}\right)\frac{1}{\zeta^2}-\frac{GMm}{2}\frac{R^2}{\zeta^4}-\frac{GMm}{2}\frac{R^4}{\zeta^6}
-\cdots.
\end{align}
The leading order consists of the sum of the centrifugal and the monopole terms and is positive for $L^2/(GMm^2)>1$ and negative for $L^2/(GMm^2)<1$. In the case $L^2/(GMm^2)=1$, the first term vanishes, and then the second term becomes leading. With any parameter set, no potential well is formed by the balance between the leading and subleading terms, but rather a potential top is formed.
This means that there only exist unstable circular orbits in the far region.

Does the absence of a stable circular orbit hold in the near region? 
As in the case where $n=3$, using the complete form of $V_4(\zeta)$, 
we can derive the energy $\mathcal{E}_0$ and the angular momentum $L_0$ of a particle in a circular orbit as 
\begin{align}
\mathcal{E}_0&
=\mathrm{sgn}(\zeta-R)\frac{GMmR^2}{2(\zeta^2-R^2)^2},
\\
L_0^2&=\mathrm{sgn}(\zeta-R)\frac{GMm^2\zeta^4}{(\zeta^2-R^2)^2},
\end{align}
where $\mathrm{sgn}(\cdot)$ denotes the sign function. 
In the region $0<\zeta<R$, there are no circular orbits because $L_0^2<0$.
On the other hand, in the region $\zeta>R$, circular orbits exist because $L_0^2>0$ but are unstable there,
\begin{align}
V_4''(\zeta)|_{L=L_0}
=-\frac{4GMm R^2}{(\zeta^2-R^2)^3}<0.
\end{align}
Consequently, we find that there are no stable circular orbits on the symmetric plane of a homogeneous circular ring source in $n=4$.

Let us consider $n=5$, where the effective potential~\eqref{eq:Vnlargezeta} reduces to
\begin{align}
V_5(\zeta)=\frac{L^2}{2\:\!m\:\!\zeta^2}-\frac{GMm}{3}\frac{1}{\zeta^3}-\frac{3GMm}{4}\frac{R^2}{\zeta^5}-\cdots.
\end{align}
The leading order turns to be the centrifugal term only, and the subleading order contains the monopole term. 
These two terms do not make a potential well but a potential top. This means that circular orbits are not stable but unstable in the far region.
Applying the same procedure as above, we obtain the energy and the angular momentum of a particle in a circular orbit as
\begin{align}
&\mathcal{E}_0=\frac{GMm}{6\pi(\zeta-R)(\zeta+R)^2}\left[\:\!
\frac{3\zeta^2+5R^2}{(\zeta-R)^2}E(z)-K(z)
\:\!\right],
\\
&L_0^2=\frac{GMm^2 \zeta^2}{3\pi (\zeta-R)(\zeta+R)^2}\left[\:\!
\frac{7\zeta^2+R^2}{(\zeta-R)^2}E(z)-K(z)
\:\!\right].
\end{align}
We find circular orbits in $R<\zeta<\infty$ because $L_0^2>0$ but not in $0<\zeta<R$ because $L_0^2<0$. However, such circular orbits in $R<\zeta<\infty$ are eventually unstable
\begin{align}
V_5''(\zeta)|_{L=L_0}
=\frac{2GMm (2\zeta^2+R^2)}{3\pi \zeta^2(\zeta-R)^2(\zeta+R)^3}\left[\:\!
K(z)-\frac{5\zeta^4+18R^2\zeta^2+R^4}{(\zeta-R)^2(2\zeta^2+R^2)}E(z)
\:\!\right]<0.
\end{align}
Consequently, we find that there are no stable circular orbits on the symmetric plane of a homogeneous circular ring source in $n=5$. As can be expected from the fact that the asymptotic form of $V_n(\zeta)$ exhibits qualitatively similar structure, the result that stable circular orbits do not exist on the symmetric plane is the same when considering $n\geq6$ (see the Appendix).

\subsection{Dynamics on the axis of symmetry $\zeta=0$}
We consider the motion of a particle constrained to the axis of symmetry of the ring, $\zeta=0$. Such a particle must have $L=0$ and feels the effective potential,
\begin{align}
V_n(\rho)=\frac{Q^2}{2m\rho^2}-\frac{GMm}{(n-2) (\rho^2+R^2)^{(n-2)/2}},
\end{align}
where we have used $r_\pm|_{\zeta=0}=\sqrt{\rho^2+R^2}$ and $F(a,b,c; 0)=1$. Throughout this subsection, we use the notation $V_n(\rho)=V_n(0, \rho)\big|_{L=0}$ for an abbreviation.
Once injected onto the axis, a particle continues to move on it, but its stability to small perturbations perpendicular to the axis is nontrivial because of the shape of the source. 
To clarify the stability, let us see the expansion of $V_n(\zeta,\rho)\big|_{L=0}$ around $\zeta=0$:
\begin{align}
V_n(\zeta,\rho)|_{L=0}=V_n(\rho)+\frac{GMm}{2(\rho^2+R^2)^{(n+2)/2}}\left(\rho^2-\frac{n-2}{2}R^2\right)\zeta^2+O(\zeta^4).
\end{align}
Whether particles remain stably bound on the axis of symmetry depend on the sign of the coefficient of $\zeta^2$ in the second term. 
If a particle moving on the axis of $\rho\geq \sqrt{(n-2)/2}R$ is perturbed perpendicularly to the axis, it continues to move stably in the vicinity of the axis. 
If a particle moving on the axis of $\rho<\sqrt{(n-2)/2}R$ is perturbed in the same way, 
it immediately moves away from the axis.

It should be noted that in the case $n=3$, the potential $V_3(\rho)$ becomes a monotonically increasing function of $\rho$, and the centrifugal force against gravity does not work. 
Therefore, no extremum point of $V_3(\rho)$ can be formed, and thus, a particle cannot stay at rest on the axis.

The following discussion will focus on the case of $n\geq 4$.
To consider particle dynamics in the region where $\rho/R\ll 1$ (i.e., near the center), we derive the expansion of $V_n(\rho)$ around $\rho=0$,
\begin{align}
V_n(\rho)=\frac{Q^2}{2m\rho^2}-\frac{GMm}{(n-2)R^{n-2}}+\frac{GMm}{2R^n}\rho^2+\cdots.
\end{align}
The balance between the centrifugal and gravitational terms can make a potential well, and therefore, a particle can stably bound in the direction along the axis near the center. However, as discussed above, since stationary orbits are unstable with respect to perturbations in the $\zeta$-direction for small $\rho$, we can finally conclude that the circular orbits near the center on the axis of symmetry is unstable. 
Note that these properties are common for $n\geq4$.

Next, we consider $V_n(\rho)$ in the asymptotic region, where $\rho/R\gg1$. 
The asymptotic expansion of $V_n(\rho)$ becomes 
\begin{align}
\label{eq:exprhoinf}
V_n(\rho)=\frac{Q^2}{2m\rho^2}-\frac{GMm}{(n-2)\rho^{n-2}}+\frac{GMmR^2}{2\rho^n}-\frac{n GMmR^4}{8\rho^{n+2}}+\cdots. 
\end{align}
The second term is the monopole term, and the terms after the third are contribution of the ring shape.

For $n=4$, the expansion~\eqref{eq:exprhoinf} takes the form,
\begin{align}
V_4(\rho)=\left(\frac{Q^2}{m}-GMm\right)\frac{1}{2\rho^2}+\frac{GMmR^2}{2\rho^4}-\frac{ GMmR^4}{2\rho^{6}}+\cdots.
\end{align}
The leading term consists of the centrifugal and monopole terms, and its sign depends on the choice of parameters. 
The leading term is positive for $Q^2/(GMm^2)> 1$, and the subleading term is also positive. 
This implies that $V_4(\rho)$ makes no potential well in the far region. 
For $Q^2/(GMm^2)=1$, since the first term vanishes, the second term becomes leading. 
The positivity of the leading term implies the nonexistence of a potential well in the far region. 
On the other hand, for $Q^2/(GMm^2)<1$, the leading is negative while the subleading is positive. Therefore, these terms can make a potential well in the far region. 
In fact, we can observe that a local minimum is formed in the far region in the case where $0<1-Q^2/(GMm^2)\ll 1$.
Consequently, we conclude that circular orbits in the far region are stable not only for the $\zeta$-direction but for the $\rho$-direction in $n=4$. 
Note that the ring shape of the source contributes to the formation of such a potential well explicitly.

For $n=5$, the expansion~\eqref{eq:exprhoinf} takes the form,
\begin{align}
V_5(\rho)=\frac{Q^2}{2m\rho^2}-\frac{GMm}{3\rho^{3}}+\frac{GMmR^2}{2\rho^5}+\cdots.
\end{align}
The leading term is the centrifugal potential and is positive. Hence, $V_5(\rho)$ makes no potential well in the far region. 
Needless to say, $V_5(\rho)$ with $Q=0$ also makes no extremum point because of the monotonicity of the Newtonian potential. 
The qualitative nature of the effective potential $V_n(\rho)$ for $n\geq 6$ is the same as that for $n=5$. In fact, 
there exist no stable circular orbits in the far region on the symmetric axis of the ring in $n\geq5$.

We clarify the whole picture of circular orbits on the axis of symmetry using the complete form of $V_n(\rho)$.
The conditions for a particle moving on a circular orbit are given by
\begin{align}
V_n'(\rho)=0, \quad V_n(\rho)=\mathcal{E},
\end{align}
where the prime denotes derivative with respect to $\rho$. Solving these equations in terms of $\mathcal{E}$ and $Q$, 
we obtain the energy and angular momentum of a particle on a circular orbit,
\begin{align}
&\mathcal{E}_0=\frac{GMm}{2(n-2)}\frac{(n-4)\rho^2-2R^2}{(\rho^2+R^2)^{n/2}},
\\
\label{eq:Q0}
&Q_0^2=\frac{GMm^2\rho^4}{(\rho^2+R^2)^{n/2}},
\end{align}
respectively. The fact that $Q_0^2>0$ for all range of $\rho$ means that 
we can find circular orbits at any point on the axis. The energy $\mathcal{E}_0$ for $n=4$ is always negative. For $n>5$, however, $\mathcal{E}_0$ is negative for $\rho<\sqrt{2/(n-4)}R$, zero for $\rho=\sqrt{2/(n-4)}R$, and positive for $\rho>\sqrt{2/(n-4)}R$. We should recall that the reference value for the energy of a particle is zero, which is consistent with the rest energy of a particle at infinity. 
The stability to small perturbations in the $\rho$-direction for a particle taking a circular orbit on the axis of symmetry requires 
\begin{align}
\label{eq:Vrho2}
V_n''(\rho)\big|_{Q=Q_0}=GMm \frac{4R^2-(n-4)\rho^2}{(\rho^2+R^2)^{(n+2)/2}}\geq0. 
\end{align}
For $n=4$, this condition always holds, and thus,
we can find a potential well at any point on the axis. 
On the other hand, for $n\geq5$, the condition~\eqref{eq:Vrho2} holds only in the range,
\begin{align}
0\leq \rho\leq \frac{2R}{\sqrt{n-4}}.
\end{align}
These results for stability are consistent with the above results of the asymptotic analysis of the stability of the circular orbits.

\begin{figure}[t]
\centering
 \includegraphics[width=12cm,clip]{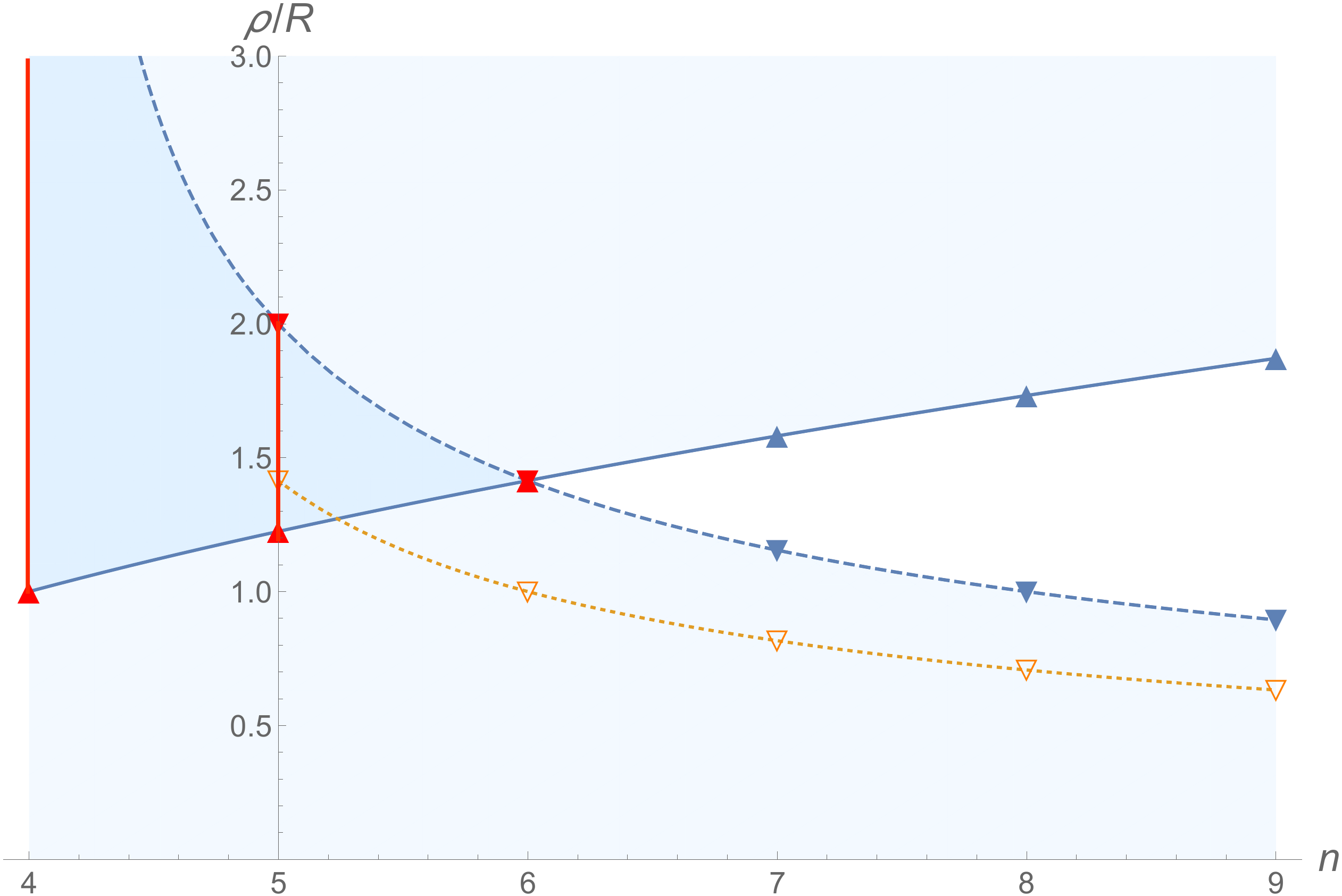}
 \caption{Range of the sequence of stable circular orbits on the axis of symmetry of a homogeneous circular ring. The blue shaded area above the solid blue line is the region where particles in circular orbits on the axis are stably bound to the axis against perturbations in the $\zeta$-direction. The blue shaded area below the dashed blue line is the region where particles in circular orbits on the axis are stably bound against perturbations in the $\rho$-direction. 
The overlap of these two shaded shows the region where stable circular orbits can exist. 
The sequences of stable circular orbits are drawn by the red solid lines. 
The orange dotted line shows zero of the energy of particles in the circular orbits. 
}
 \label{fig:stability}
\end{figure}

Finally, let us summarize the results for stable circular orbits on the axis of symmetry. 
For $n=3$, no stable circular orbits exist on the axis 
since the direction of motion is 1D, 
and the gravitational force is attractive. For $n\geq4$, the sequences of stable circular orbits are visualized in Fig.~\ref{fig:stability}. 
For $n=4$, we obtain the sequence of stable circular orbits in the range, 
\begin{align}
R\leq \rho<\infty \quad (n=4), 
\end{align}
which is shown by a red half line in Fig.~\ref{fig:stability}. Within the range, we have $\mathcal{E}_0<0$. 
Note that the sequence terminates at $\rho=R$. The circular orbit here is the ISCO with $\mathcal{E}_0=-GMm/(8R^2)$ and $Q_0^2=GMm^2/4$. In the range where $0\leq \rho<R$, circular orbits exist but are unstable. 
For $n=5$, we obtain the sequence of stable circular orbits in the range,
\begin{align}
\frac{\sqrt{6}}{2}\leq \frac{\rho}{R}\leq 2 \quad (n=5),
\end{align}
which is shown by a red segment in Fig.~\ref{fig:stability}. 
As seen in the asymptotic analysis, while there is no mechanism to have a local minimum of the effective potential in the far region, a local minimum can be formed near the center by the balance between a centrifugal barrier and gravitational force. We can understand $\rho=2R$, the radius of the outermost stable circular orbit, 
as a switching point between these two mechanisms. 
However, the sequence terminates at $\rho=\sqrt{6}R/2$, the radius of the ISCO, because the bound of particles in circular orbits becomes unstable against perturbations in the $\zeta$-direction here. 
It is worth noting that the energy of a particle in a stable circular orbit is not positive in the range 
$\sqrt{6}/2\leq \rho/R\leq \sqrt{2}$. On the other hand, the energy is positive in the range 
$\sqrt{2}<\rho/R<2$. 
For $n=6$, there exists a marginally stable circular orbit only at the radius,
\begin{align}
\rho=\sqrt{2}R \quad (n=6),
\end{align}
which is shown by a red point in Fig.~\ref{fig:stability}. 
The energy and angular momentum of a particle to stay here is 
$\mathcal{E}_0=GMm/(108 R^4)$ and 
$Q_0^2=4GMm^2/(27R^2)$, respectively. 
For $n\geq 7$, we have no stable circular orbits on the axis of symmetry.

\section{Summary and discussions}
\label{sec:4}

In this paper, we have considered the Newtonian gravitational potential sourced by a homogeneous circular ring in $\mathbb{E}^n$ and have seen that the potential in even-dimensional space is greatly simplified into a form with finite terms. We can expect that the parity of the potential must be shared by black ring solutions in general dimensions; for example, the spacetime metrics of even spatial dimensions would have a simpler structure than those of odd. A new exact solution of the black rings may be found not from 5D space but from 6D space earlier.

In the main part, we have considered the dynamics of freely falling massive particles in the Newtonian potential, in particular stable circular orbits on the symmetric plane, which is the plane containing the ring, and on the axis of symmetry. One of the results is that particles cannot move in a circular orbit stably 
on the symmetric plane in higher-dimensional space. We can understand it as the disappearance of stable equilibrium between the centrifugal and gravitational forces caused by the increase of the inverse power of gravity as increasing spatial dimensions. In 3D space, however, stable circular orbits exist from infinity to a certain radius larger than the radius of the ring. This lower bound corresponds to the ISCO, which is familiar in black hole spacetimes, but it is not the only case that appears.

It is worth noting that 
the phenomenon of the ISCO we have seen in 3D is not due to ring properties such as symmetry. 
One of the mechanisms of the ISCO formation is the degeneracy of the radii of stable and unstable circular orbits. It occurs, for example, when the gravitational potential dominates both at infinity and at a certain location such as the source position, and the centrifugal potential becomes dominant between them. 
This is exactly the case with the behavior of the effective potential in 3D space.
A homogeneous ring is not the only source of such a situation, and it is worthwhile to consider various sources that form the ISCO by a similar mechanism.

We have also clarified that particles constrained on the axis of symmetry of the ring can only move linearly in 3D space, while they are allowed to take circular orbits in the direction of extra dimensions in higher-dimensional spaces. In 4D space, in particular, a sequence of stable circular orbits exists from infinity to the ISCO radius, which is nonzero. In 5D space, such a sequence exists from the radius of the outermost stable circular orbit to the nonzero ISCO radius. On the other hand, in spaces higher than 6D, only unstable circular orbits appear. These are the results of the dimensional dependence of the gravitational field created by the ring.

Our results reveal a part of the nature of timelike geodesics of higher-dimensional black ring solutions through the most fundamental particle orbits: stable circular orbits in the Newtonian gravity. Based on the blackfold approach, we can approximate the gravitational field in the vicinity of a black ring with a bent black string. Since the dynamics of particles in the vicinity of an $(n+1)$-dimensional black string are equal to those of an $n$-dimensional black hole, we can predict its behavior relatively easily. 
The full picture will become clearer when the exact solution of the higher-dimensional black ring is discovered, and the behavior of the timelike geodesics, including the vicinity of the horizon, is clarified. Clarification of the entire sequences of stable circular orbits in the region away from the symmetry plane and axis of symmetry is an issue for the future. A branch of the sequences may extend towards the ring. While the equation of motion is integrable in 4D case, it remains unsolved whether or not it is also integrable in the case of more than 5D. With the change in dimensions, this system may even manifest chaotic instead of integrable.

\begin{acknowledgments}
The author are grateful to S.~Kinoshita for useful comments. 
This work was supported by Grant-in-Aid for Early-Career Scientists~(JSPS KAKENHI Grant No.~JP19K14715). 
\end{acknowledgments}
\appendix

\section{Circular orbits on the symmetric plane}
\label{sec:A}
The conditions satisfied by a particle moving in a circular orbit on the symmetric plane are
\begin{align}
V_n'(\zeta)=0, \quad V_n(\zeta)=\mathcal{E},
\end{align}
where the explicit form of $V_n(\zeta)$ is given in Eq.~\eqref{eq:Vncomplete}.
Solving these in terms of $\mathcal{E}$ and $L$, we obtain 
\begin{align}
\mathcal{E}_0&=\frac{GMm}{2(\zeta+R)^{n-1}}\left[\:\!
\frac{
(n-4)\zeta-2R}{n-2}\:\!
F\left(\frac12, \frac{n-2}{2}, 1, z\right)
+\frac{R\:\!\zeta(\zeta-R)}{(\zeta+R)^2}\:\!F\left(
\frac32, \frac{n}{2}, 2, z
\right)
\:\!\right],
\\
L_0^2&=\frac{GMm^2\zeta^3}{(\zeta+R)^{n-1}}\left[\:\!
 F\left(
\frac12, \frac{n-2}{2}, 1, z
\right)+\frac{R(\zeta-R)}{(\zeta+R)^2}\:\! F\left(
\frac32, \frac{n}{2}, 2, z
\right)
\:\!\right],
\end{align}
respectively. According to the results of numerical analysis of $L_0^2$, 
there seem to be no circular orbits in $0\leq \zeta<R$ because $L_0^2<0$. 
On the other hand, there seem to be circular orbits in $\zeta>R$ because $L_0^2>0$. 
The second derivative of $V_n(\zeta)|_{L=L_0}$ evaluated at $\zeta>R$ is 
\begin{align}
V_n''(\zeta)\big|_{L=L_0}
=\frac{GMm}{2\zeta (\zeta+R)^{n+2}}
\bigg[\:\!u(\zeta)\:\! F\left(\frac{3}{2}, \frac{n}{2}, 2, z\right)
+v(\zeta)\:\!F\left(\frac{5}{2}, \frac{n}{2}, 2, z
\right)
\:\!\bigg],
\end{align}
where
\begin{align}
u(\zeta)&=(n-4) \zeta^3+2\left(2n^2-13n+17\right)R\zeta^2+(8-7n)R^2\zeta-6R^3,
\\
v(\zeta)&=3(\zeta-R)^2\left[\:\!
2R-(n-4)\zeta
\:\!\right].
\end{align}
The numerical analysis shows that $V_n''(\zeta)\big|_{L=L_0}<0$ for $n\geq4$, which means that the circular orbits are unstable.

\end{document}